\documentclass[twocolumn,aps,prl,superscriptaddress,showpacs]{revtex4-1}
\usepackage[utf8]{inputenc}
\usepackage{color}
\usepackage{amsmath}
\usepackage{amssymb}
\usepackage{graphicx}
\usepackage{esint}
\usepackage[unicode=true,
 bookmarks=false,
 breaklinks=true,pdfborder={0 0 0},backref=false,colorlinks=true]
 {hyperref}
\usepackage{xcolor}
\makeatletter

\usepackage{epstopdf}\usepackage{subfigure}
\usepackage{float}
\usepackage{xcolor}
\usepackage{amsfonts}
\usepackage{bm}
\usepackage{subfigure}

\makeatother

\begin{document}

\title{Predicting topological materials: symmetry-based indicator theories and beyond}

\author{Tiantian Zhang}

\author{Shuichi Murakami}
\affiliation{Department of Physics, Tokyo Institute of Technology, Ookayama, Meguro-ku, Tokyo 152-8551, Japan}
\affiliation{Tokodai Institute for Element Strategy, Tokyo Institute of Technology, Nagatsuta, Midori-ku, Yokohama, Kanagawa 226-8503, Japan}

\begin{abstract}
Though symmetry-based indicators formulae are powerful in diagnosing topological states with a gapped band structure at/between any high-symmetry points, it fails in diagnosing topological degeneracies when the compatibility condition is violated. In such cases, we can only obtain information of whether there is a band degeneracy at some high-symmetry points or along some high-symmetry lines by the compatibility condition. 
Under the framework of symmetry-based indicator theories, we proposed an algorithm to diagnose the topological band crossings in the compatibility condition-violating systems to obtain the whole topological information, by using the symmetry-based indicator formulae of their subgroups. 
In this paper, we reinterpret the algorithm in a simpler way with two material examples preserving different topological states in spinless systems with time-reversal symmetry, discuss the limitation of the symmetry-based indicator theories, and make further discussions on the algorithm applying in spinful systems with time-reversal symmetry.

\end{abstract}

\maketitle

\section{Introduction}

\begin{center}
\begin{figure}
\includegraphics[scale=0.8]{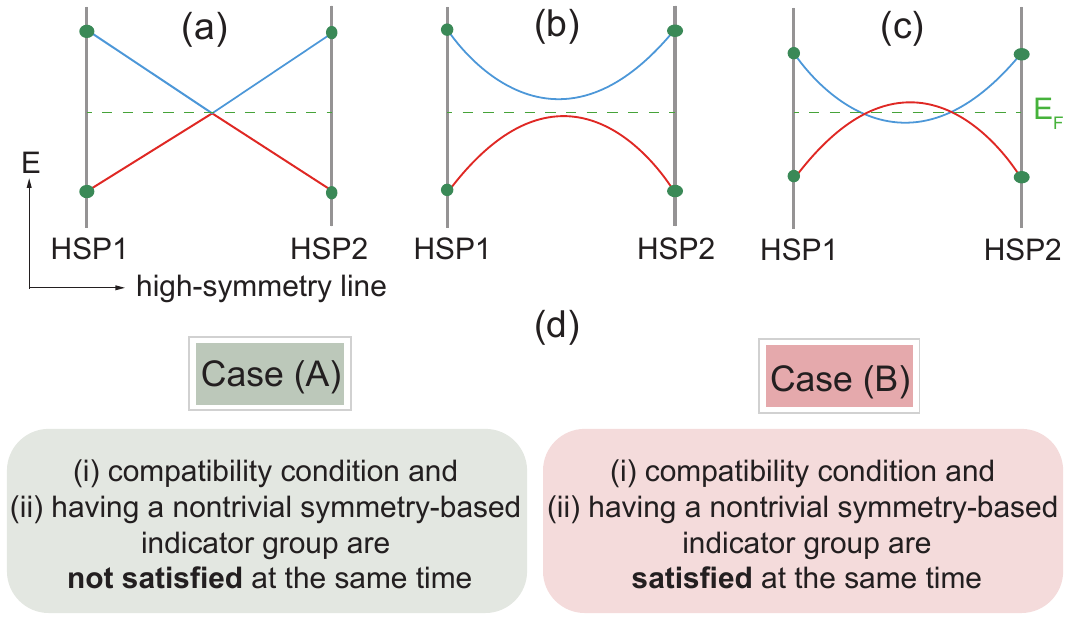}\caption{Three different kinds of band structure between two high-symmetry points, where the red and blue line represents the occupied and the unoccupied band, respectively. HSP1 and HSP2 are two different high-symmetry points in the Brillouin zone. (a) Band structure violating the compatibility condition. (b-c) Two different band structures satisfying the compatibility condition. (d) Two cases where we cannot (Case (A)) and can (Case (B)) use the symmetry-based indicator formulae to diagnose topological information. \label{fig:CC}}
\end{figure}
\end{center}

Symmetry-based indicator theories \cite{po2017symmetry,bradlyn2017TQC,song2018quantitative,song2018diagnosis,Khalaf2018symmetry} greatly decrease the calculation to diagnose topological phases of a material, and tremendous progress has been made in effectively diagnosing topological materials in the past few years \cite{zhang2019catalogue,vergniory2019complete,tang2019comprehensive}, such as symmetry-based indicator formulae for topological insulators\cite{song2018quantitative,Khalaf2018symmetry}, topological crystalline insulators\cite{song2018quantitative,Khalaf2018symmetry}, and topological semimetals\cite{song2018diagnosis}. 
However, on one hand, two conditions should be met before the application of those formulae: (i) the number of each irreducible representation at high-symmetry points (HSPs) should satisfy the compatibility condition, i.e., the band structure either has no band crossing at/between any of those HSPs or has band crossing along high-symmetry lines (HSLs) can be gapped out without making an additional band inversion at any HSPs, as shown in Fig.~\ref{fig:CC} (b-c); (ii) The system should preserve a nontrivial symmetry-based indicator group, which means that symmetry-based indicator formulae are restricted in limited space groups. 
On the other hand, as presented in our previous work \cite{zhang2019catalogue}, more than 62\% (98\%) materials are diagnosed to be a topological semimetal in the spinful (spinless) case. 
Among those topological semimetals, which are the majority of the diagnosed topological materials, 100\% of them in the spinful case and 99\% in the spinless case violate the condition (i), which means that we cannot use the symmetry-based indicator formulae to get further topological information with a simple calculation. 

However, according to the algorithm proposed in our previous work \cite{zhang2020diagnosis}, most of the topological semimetals can be further diagnosed by the symmetry-based indicator of their subgroups to obtain the whole topological information. Such algorithm requires one to ignore some crystalline symmetries of the system, so as to obtain a subgroup which meets both of Condition (i) and Condition (ii) at the same time. 
Yet, still a minority of them cannot be diagnosed by the symmetry-based indicators under this algorithm, but only by the compatibility condition, and we will also discuss such cases in detail. Thus, such diagnosing algorithm is under the symmetry-based indicator theories, but also beyond them. 

In this paper, we will first give a brief overview of the Condition (i) and Condition (ii), which are two basic conditions before using the symmetry-based indicator formulae to diagnose topological phases. Second, two cases in different topological phases are analyzed by the symmetry-based indicator theories and beyond, and then followed by the restriction of the symmetry-based indicator theories. In the end, similar discussions for the algorithm applied in spinful systems with time-reversal symmetry ($\mathcal{T}$) are outlined in the last section.


\section{Compatibility condition and symmetry-based indicator group}
Compatibility conditions restrict the number of symmetry data, i.e., a set of irreducible representations for the occupied bands at HSPs in the Brillouin zone (BZ). Since such condition only involves the symmetry data at HSPs, it can only tell the information of band inversion at HSPs rather than along HSLs, compared to the band structure of an atomic insulator \cite{po2017symmetry,bradlyn2017TQC}. 
For example, band structure with a band inversion at a HSP in Fig.~\ref{fig:CC} (a) doesn't satisfy the compatibility condition, while band structure with a band inversion along a HSL in Fig.~\ref{fig:CC} (c) will satisfy the compatibility condition like Fig.~\ref{fig:CC} (b), since such a band inversion can be gradually gapped out through by a perturbation without making a new band inversion at any HSPs. 
Thus, a topological semimetal like SrSi$_2$ \cite{huang2016new} with a band inversion along a HSL, like Fig.~\ref{fig:CC} (c), will not be considered here due to the satisfaction of the compatibility condition and diagnosed as a trivial insulator by the symmetry-based indicator theories. 
Namely, it is a shortcoming of the compatibility condition that a band structure like Fig.~\ref{fig:CC} (c) is diagnosed also as a ``gapped" phase. 
Thus, if the symmetry data of a system satisfies the compatibility condition, there will be no band crossings at any HSPs or along any HSLs, which corresponds to a``gapped" band structure. Likewise, if the symmetry data of a system does not satisfy the compatibility condition, there will be a band crossing at some HSPs or along some HSLs, but the Condition (i) can still be satisfied if we ignore some crystalline symmetries of the system. 

Symmetry-based indicators are powerful in predicting topological materials, because we can tell the topological information of a material by just calculating the symmetry data at up to 8 HSPs \cite{po2017symmetry,bradlyn2017TQC,song2018quantitative,song2018diagnosis,Khalaf2018symmetry,zhang2019catalogue,vergniory2019complete,tang2019comprehensive}. However, systems with certain space groups have a trivial symmetry-based indicator group \cite{po2017symmetry}, which means that the symmetry data only at HSPs is not enough to diagnose the topological states of such systems, but wavefunctions at other $k$ points need to be calculated. In some cases, ignoring some crystalline symmetries may give rise to a nontrivial symmetry-based indicator group, which will still help us to diagnose the topological phases.

As we discussed in the $introduction$ part, 
symmetry-based indicator formulae cannot be used when either/both of Condition (i) and (ii) are violated, and we call such situation $Case\ (A)$. In order to proceed to use the symmetry-based indicator formulae for further diagnosis, we need to ignore some crystalline symmetries to make $Case\ (A)$ to $Case\ (B)$, where both of the conditions are satisfied, as shown in Fig.~\ref{fig:CC} (d). 

In the following, we will give two examples with Case (A) in the AI class system (spinless system with $\mathcal{T}$) \cite{schnyder2008classification,chiu2016classification,ryu2010topological}, and show how to reach Case (B) and to use symmetry-based indicator formulae to obtain the topological information afterwards. One example is Weyl phonons in a noncentrosymmetric material ZrPdSn, which satisfies Condition (i) but breaks Condition (ii); the other one is node-ring phonons in a centrosymmetric material Na$_{3}$BrO, which break both Condition (i) and (ii).

\begin{figure*}
\includegraphics[scale=0.92]{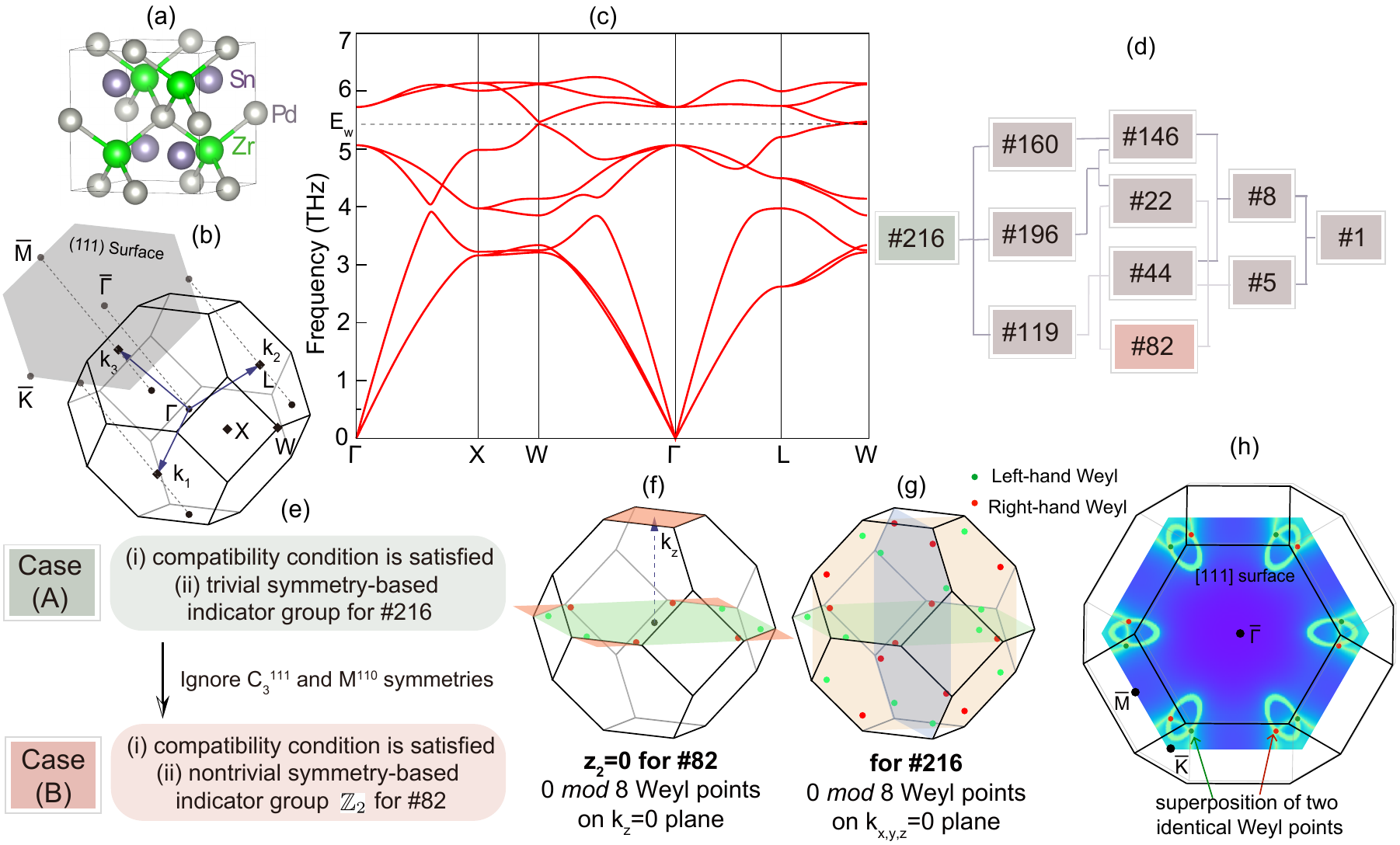}\caption{(a) Crystal structure, (b) Brillouin zone and (c) phonon band structure of ZrPdSn. (d) Subgroups of \#216. (e) Case (A) for \#216 and Case (B) for \#82 after ignoring $C_{3}^{111}$ symmetry and $M^{110}$ symmetry. (f) One possible configuration of $z_2=0$ for \#82. Red and green dots represent Weyl points with different chiralities. (g) Distribution of Weyl points for ZrPdSn with \#216. (h) Surface arcs for 24 Weyl points on the [111] surface, where every pair of two Weyl points with opposite chirality will be projected onto one point.}
\label{fig:ZrPdSn}
\end{figure*}

\begin{table}[]
\begin{tabular}{cc}
\hline
\multicolumn{2}{c}{ZrPdSn}                                        \\ \hline \hline
\multicolumn{1}{c|}{k points} & irreducible representations       \\ \hline
\multicolumn{1}{c|}{$\Gamma$} & 2$\Gamma_{4}$                     \\ \hline
\multicolumn{1}{c|}{L}        & 2L$_{1}$+2L$_{3}$                 \\ \hline
\multicolumn{1}{c|}{W}        & 2W$_{1}$+W$_{2}$+2W$_{3}$+W$_{4}$ \\ \hline
\multicolumn{1}{c|}{X}        & X$_{1}$+X$_{3}$+2X$_{5}$          \\ \hline
\end{tabular}
\caption{Irreducible representations for the lowest 6 bands of ZrPdSn at each high-symmetry momentum. The symmetry data for \#216 satisfies the compatibility condition. }
\label{table:216}
\end{table}

\section{Diagnosis process for Weyl phonons in ZrPdSn}
ZrPdSn is a narrow-band-gap semiconductor with half-Heusler crystal structure of space group $F\bar{4}3m$ (\#216)\cite{ZrPdSn}, as shown in Fig.~\ref{fig:ZrPdSn} (a). 
The compatibility condition is satisfied for the lowest 6 phonon bands according to the symmetry data at HSPs shown in Tab.~\ref{table:216}. However, the phonon spectra in Fig.~\ref{fig:ZrPdSn} (c) shows that the 6$^{th}$ band and the 7$^{th}$ band are very close to each other at around 5.4 THz, and such kind of band structure may bring out Weyl points composed of the 6$^{th}$ band and the 7$^{th}$ band at some momenta. 
In order to verify our assumption, we would like to use the symmetry-based indicator to have a fast  but accurate diagnosis for the existence of Weyl phonons in ZrPdSn. However, ZrPdSn belongs to Case (A) since the Condition (ii) is not satisfied due to the trivial symmetry-based indicator group of \#216 \cite{po2017symmetry}. 
Moreover, the symmetry data of \#216 already satisfies the Condition (i), which will also be satisfied for any of the subgroups. 
Thus, all we need to do is to ignore some symmetries of ZrPdSn, and search for the maximal subgroup(s) of \#216 having a nontrivial symmetry-based indicator group. 
Figure.~\ref{fig:ZrPdSn} (d) shows all the subgroups of \#216, but only \#82 has a nontrivial symmetry-based indicator group $\mathbb{Z}_{2}$ \cite{po2017symmetry}. Thus, ignoring $C_{3}^{111}$ and $M^{110}$ symmetries will turn ZrPdSn from the Case (A) to Case (B), where we can use the symmetry-based indicator formulae of the subgroup \#82 to diagnose the Weyl phonons, as shown in Fig.~\ref{fig:ZrPdSn} (e). 

By calculating the $\mathbb{Z}_2$-indicator of \#82, we get $z_2=0$, which indicates that there are 0 $mod$ 8 Weyl points on $k_{z}$=0 plane. Figure~\ref{fig:ZrPdSn} (f) shows one of the possible configurations of $z_2=0$, which has 8 Weyl points on the $k_{z}$=0 plane \cite{song2018diagnosis}. Red and green dots represent Weyl points with different chiralities. Thus, after reconsidering the ignored $C_{3}^{111}$ symmetry and $M^{110}$ symmetry, we can obtain a total of 24 Weyl points for \#216, located on the $k_{x}$=0, $k_{y}$=0, and $k_{z}$=0 planes, separately. A numerical calculation by density-function theory (DFT) confirms the result obtained by the symmetry-based indicator formula of the subgroup \#82. Figure~\ref{fig:ZrPdSn} (h) show the surface arcs of those 24 Weyl points along [111] direction, and every pair of two Weyl points with opposite chirality will be projected onto one point, leading to three closed surface arcs on the [111] surface. 
We note that those 24 Weyl points composed of the 6$^{th}$ band and 7$^{th}$ are ideal ones, in the sense that all of them are related by symmetries, and thus they have an equal energy and clean surface arcs.

 \section{Diagnosis process for node-ring phonons in Na$_{3}$BrO}
\begin{table}[]
\begin{tabular}{cc}
\hline
\multicolumn{2}{c}{Na$_{3}$BrO}                                                    \\ \hline
\multicolumn{1}{c|}{k points} & irreducible representations                   \\ \hline
\multicolumn{1}{c|}{$\Gamma$} & 2$\Gamma_{4-}$+$\Gamma_{5-}$                 \\ \hline
\multicolumn{1}{c|}{M}        & M$_{2+}$+M$_{2-}$+M$_{3-}$+M$_{5+}$+2M$_{5-}$ \\ \hline
\multicolumn{1}{c|}{R}        & 2R$_{4-}$+R$_{5-}$                            \\ \hline
\multicolumn{1}{c|}{X}        & X$_{1+}$+X$_{2+}$+X$_{3-}$+X$_{5+}$+2X$_{5-}$ \\ \hline
\end{tabular}
\caption{Irreducible representations for the lowest 9 phonon bands of Na$_{3}$BrO at each high-symmetry momentum. The symmetry data for \#221 breaks the compatibility condition.}
\label{table:221}
\end{table}

\begin{center}
\begin{figure*}
\includegraphics[scale=0.8]{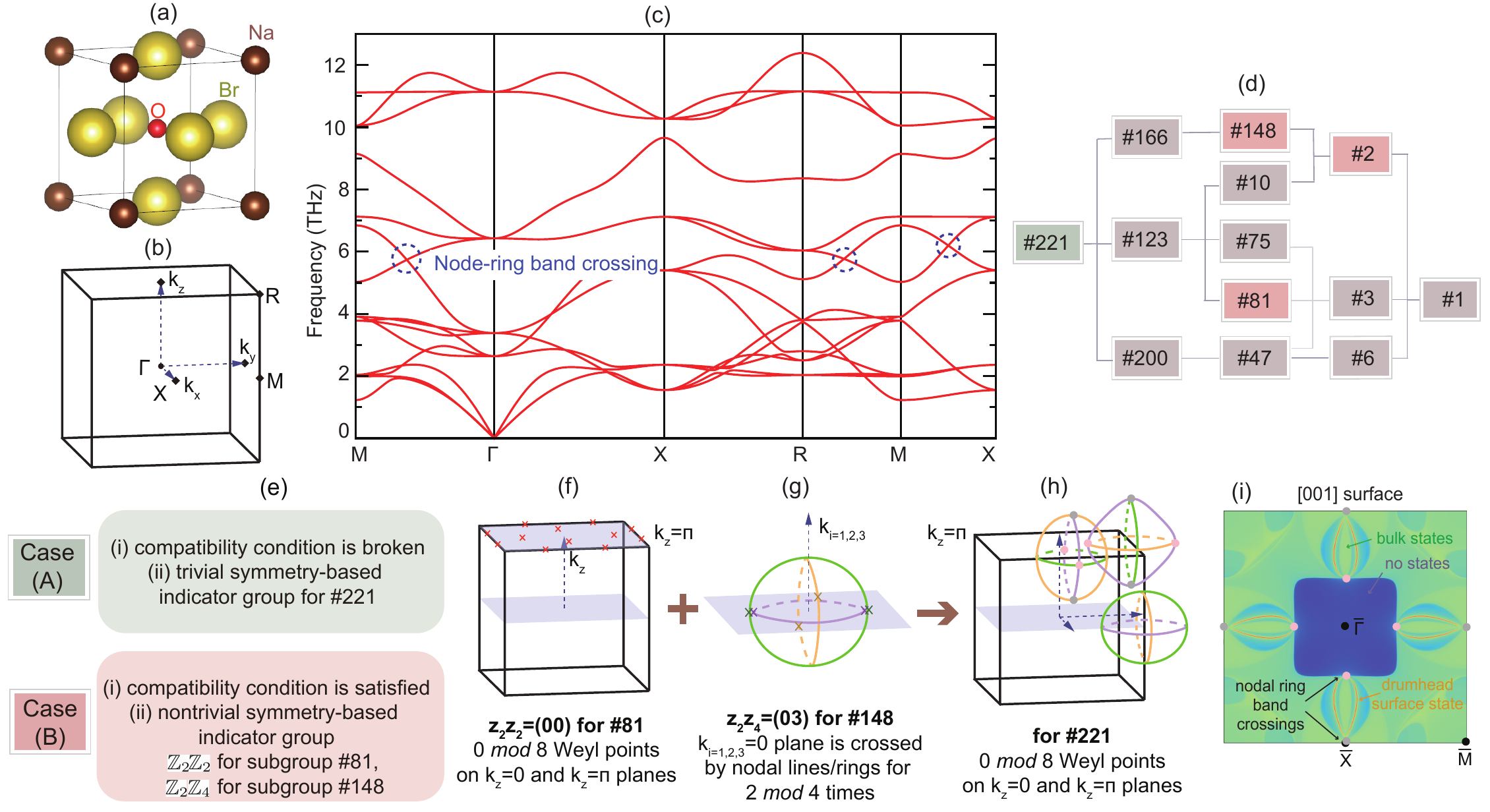}
\caption{(a) Crystal structure, (b) Brillouin zone and (c) phonon spectra for Na$_{3}$BrO.
(d) Subgroups for \#221. (e) Case (A) for \#221 and Case (B) for both subgroup \#81 and \#146 after ignoring different symmetries. (f) One of the possible configurations for subgroup \#81 with $z_{2}z_{2}$=(00). (g) One of the possible configurations for subgroup \#148 with $z_{2}z_{4}$=(03). (h) Nodal ring distribution for Na$_{3}$BrO by both symmetry-based indicator analysis and first-principle calculation. (i) Surface states for node-ring phonons in Na$_{3}$BrO on [001] surface at 5.76 THz. 
\label{fig:NodalRing}}
\end{figure*}
\end{center}
Na$_{3}$BrO is also a narrow-gap insulator but with an anti-perovskite oxide structure of space group Pm-3m (\#221) \cite{na3obr}, as shown in Fig.~\ref{fig:NodalRing}(a). From the phonon spectra in Fig.~\ref{fig:NodalRing}(c), one may notice that there are several band crossing along $\Gamma$-M, R-M and M-X directions around 5.8THz, crossed by the 9$^{th}$ band and 10$^{th}$ band. 
Those band crossings along HSLs indicate the violation of Condition (i) for the lowest 9 bands, which is also confirmed by the symmetry data of Na$_{3}$BrO shown in Tab.~\ref{table:221}. 
Furthermore, \#221 has a trivial symmetry-based indicator group \cite{po2017symmetry}, which also violates Condition (ii). 
Thus, we need to ignore some symmetries to find maximal subgroup(s) which satisfy both of the conditions, and proceed with the symmetry-based indicator analysis. By following the subgroup list shown in Fig.~\ref{fig:NodalRing}(d), we found two maximal subgroups of \#221 for Na$_{3}$BrO, i.e., \#148 (R-3) and \#81 (I-4), and neither of them is a subgroup of the other one. 
For \#148, all the mirror symmetries and $C_{4}$ symmetries are ignored, while for \#81, all the mirror symmetries, inversion symmetry, $C_{3}^{111}$, all the $C_{4}$ symmetries, all the $C_{2}$ symmetries except for the $C_{2z}$ symmetry are ignored. 
For subgroups, $e.g.$, \#148, after ignoring all the mirror symmetries and $C_{4}$  symmetries, each irrep from \#221 can be mapped to \#148. Then, we need to use the compatibility condition of the subgroup \#148, which is different from \#221, to check whether the mapped irreps satisfy the compatibility condition or not. 
Both of the subgroups satisfy the (i) compatibility condition and (ii) nontrivial symmetry-based indicator group condition, and can alter Na$_{3}$BrO from Case (A) to Case (B), as shown in Fig.~\ref{fig:NodalRing}(e). 


For subgroup \#81, S$_{4}$ symmetry is the only generator. This group has a nontrivial indicator group $\mathbb{Z}_{2}\mathbb{Z}_{2}$ \cite{po2017symmetry}, and it takes a value of $z_{2}z_{2}$=(00) for the symmetry data of Na$_{3}$BrO. Such zero indicators of \#81 tell us that there are 0 $mod$ 8 band crossings both on $k_{z}=0$ plane and on $k_{z}=\pi$ plane, and the number of band crossings on those two planes can be different. Fig.~\ref{fig:NodalRing}(f) shows one possible configuration of the band crossings with $z_{2}z_{2}$=(00), i.e., 8 band crossings on the $k_{z}=\pi$ plane and 0 band crossings on the $k_{z}$=0 plane, where band crossings are marked by red crosses. Since both the $\mathcal{T}$ and inversion symmetry are preserved in Na$_{3}$BrO, the band crossings may belong to nodal lines/rings instead of isolated Dirac points. Thus, only with the information given by the indicator of subgroup \#81 we cannot get the whole topological configuration for Na$_{3}$BrO. Thus, we need to proceed to analyze the topological information given by the indicators of the other subgroup \#148. 

An indicator group for \#148 is $\mathbb{Z}_{2}\mathbb{Z}_{4}$, which is actually equivalent to the indicator group of $\mathbb{Z}_{2}\mathbb{Z}_{2}\mathbb{Z}_{2}\mathbb{Z}_{4}$ for \#2, with the first three indicators $\mathbb{Z}_{2}$ having the same value. In addition, for space groups with inversion symmetry, subgroup \#2 is always a choice to turn the system from Case (A) to Case (B), because the compatibility condition for \#2 is alway satisfied for any symmetry data. Symmetry data of \#148 (\#2) gives rise to indicators of $z_{2}z_{4}$=(03) ($z_{2}z_{2}z_{2}z_{4}$=(0003)), which implies that the $k_{i=1,2,3}$=0 plane will be crossed by nodal lines/rings for 2 $mod$ 4 times. Fig.~\ref{fig:NodalRing}(g) shows one of the possible configurations of $z_{2}z_{4}$=(03), i.e., 3 nodal rings cross $k_{i}$ =0 plane for 6 times, with two of the nodal rings perpendicular to the $k_{i}$ =0 plane and one of them is almost lie in the plane. They do not necessarily lie on the $k_{i}$ =0 planes since there is no mirror symmetry for subgroup \#148. Since the symmetry-based indicator groups are Abelian groups, they should satisfy the sum rule. Thus, the indicators of $z_{2}z_{2}z_{2}z_{4}$=(0003) for \#2 can be understood as a sum of three separated indicators:  (1101)+(0111)+(1011)=(0003), which means that there are 1 $mod$ 2 nodal rings located at ($\pi$,$\pi$,0), (0,$\pi$,$\pi$) and ($\pi$,0,$\pi$), separately. Thus, one of the possible configurations for the topological degeneracies is 3 nodal rings like Fig.~\ref{fig:NodalRing}(g) located at ($\pi$,$\pi$,0), (0,$\pi$,$\pi$) and ($\pi$,0,$\pi$), separately.

Combining the indicators of two subgroups, we can get a possible configuration of topological band crossings in Na$_{3}$BrO, which is 9 nodal rings in the Brillouin zone, and each three of them locate around ($\pi$,$\pi$,0), (0,$\pi$,$\pi$) and ($\pi$,0,$\pi$), separately. In order to make the nodal rings in closed loops, we show the distributions of those band crossings around ($\pi$,$\pi$,$\pi$) crossing 8 BZs, as shown in Fig.~\ref{fig:NodalRing} (h). 
Since the existence of mirror symmetry for \#221, all of those 9 nodal rings will lie on the $k_{i=1,2,3}$ =0 ($k_{i=x,y,z}$ =0) mirror-preserved planes, which is also confirmed by the DFT calculation. 
Figure ~\ref{fig:NodalRing} (i) shows the drumhead surface states of those nodal rings along [001] direction at 5.76 THz, where the pink and grey dots are the projection of nodal rings and also marked in Fig. ~\ref{fig:NodalRing} (h). All of them correspond to the band crossing along X-M direction in Fig. ~\ref{fig:NodalRing} (c). Since the projection of the grey dots along the surface boundary contains four node-ring band crossings, there will be four drumhead surface states coming out form those points, as shown by the light orange lines in Fig. ~\ref{fig:NodalRing} (i). Each projection of pink dot contain two node-ring band crossings, thus there will be two drumhead surface states coming out from each pink point. 
We note that the position and number of surface states will change with different energies, since node-ring band crossings not related by symmetries will have different energies.

\section{Topological states that cannot be diagnosed by symmetry-based indicators}

\begin{center}
\begin{figure}
\includegraphics[scale=0.8]{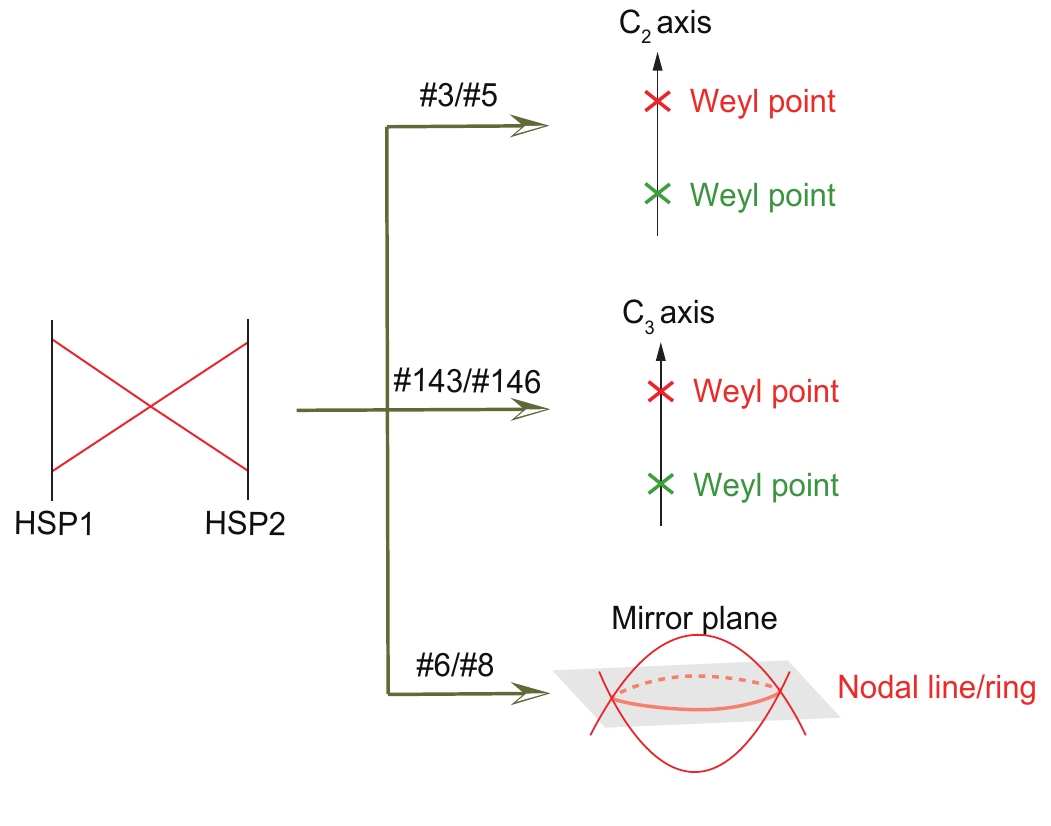}\caption{Three cases where the topological degeneracies can be only diagnosed by compatibility condition, but cannot by the symmetry-based indicator. (a) System only with C$_{2}$ symmetry. (b) System only with C$_{3}$ symmetry. (c) System with a single mirror symmetry. 
\label{fig:3SGs} }
\end{figure}
\end{center}

In the previous sections, we introduced the algorithm to use the symmetry-based indicator formulae of subgroups to diagnose topological states even when the system is in Case (A). 
However, not all the systems have a subgroup which can turn the system from Case (A) to Case (B), i.e., not all the topological band crossings of a system can be diagnosed by the indicators of its subgroups. In such cases, the subgroup search will end up with \#1 with no subgroups satisfying both Conditions (i) and (ii). 
In this paper, we will introduce three simplest cases where topological degeneracies can be only diagnosed by the compatibility condition, but not the symmetry-based indicators. They are systems only with C$_{2}$ symmetry (e.g., \#3 and \#5), systems only with C$_{3}$ symmetry (e.g., \#143 and \#146) and systems with a single mirror symmetry (e.g., \#6 and \#8). We note that topological information for those cases are still easy to be obtained since crystalline symmetries are very simple. Other cases with more complex symmetries can be treated as a combination of those three cases. 
If a system only with C$_{2}$/C$_{3}$ symmetry but still violates the compatibility condition, it is easy to know that there is a pair of Weyl points with opposite chirality locating along the C$_{2}$/C$_{3}$ axis, as shown in Fig.~\ref{fig:3SGs}. If a system only has a mirror symmetry and the compatibility condition is violated, there will be node-line/ring band crossings locating on the mirror plane, as shown in Fig.~\ref{fig:3SGs}.

\begin{center}
\begin{figure}
\includegraphics[scale=0.8]{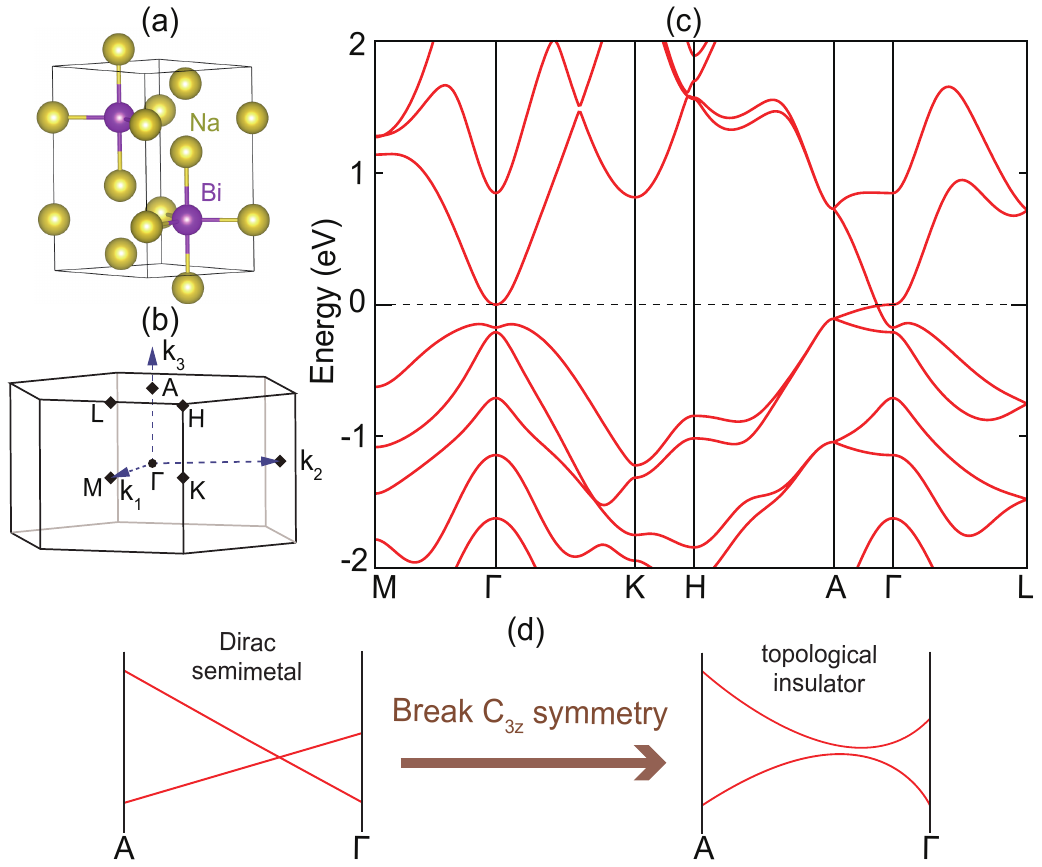}\caption{(a) Crystal structure, (b) Brillouin zone and (c) spinful electric band structure for Na$_{3}$Bi. (d) Na$_{3}$Bi will become a topological insulator with a perturbation breaking C$_{3z}$ symmetry. \label{fig:Na3Bi} }
\end{figure}
\end{center}
%

\section{Diagnosing topological states by symmetry-based indicator in class AII systems}

The algorithm of diagnosing topological states by the indicator of subgroups can also be used in other symmetry classes with minor changes, such as class AII systems, i.e., spinful systems with $\mathcal{T}$. In class AII systems, the compatibility condition can help us to diagnose topological semimetals with band crossings located at HSPs or along HSLs, but symmetry-based indicators are used to diagnose gapped states like trivial insulators, topological insulators (TIs) and topological crystalline insulators (TCIs). Thus, by using the indicators of subgroups, one can obtain the topological phase transition from a topological semimetal to trivial insulator/TI/TCI when certain crystalline symmetries are broken by perturbations.  

For example, Na$_{3}$Bi is a Dirac semimetal with space group \#194. There are two Dirac points located along A-$\Gamma$-A HSLs and protected by $\mathcal{T}$ and C$_{3z}$ rotation symmetry, which also indicates violation of the compatibility condition. In order to use the symmetry-based indicator of subgroups to diagnose the topological gapped states for Na$_{3}$Bi with certain perturbation, we need to consider the maximal subgroup of Na$_{3}$Bi without C$_{3z}$ symmetry, which is \#12. Indicators for subgroup \#12 are $z_{2}z_{2}z_{2}z_{4}$=(0003) indicating a strong TI phase, so a phase transition from Dirac semimetal to TI will be obtained by breaking C$_{3z}$ symmetry in Na$_{3}$Bi. 

We note that both Na$_{3}$Bi with \#12 and Na$_{3}$BrO with \#2 have the same indicator of $z_{2}z_{2}z_{2}z_{4}$=(0003), but correspond to different topological states, i.e., the former one is a strong TI while the latter one is nodal ring. Such difference comes from different systems of those two materials. 
For Na$_{3}$Bi, the system belongs to a spinful one with $\mathcal{T}$, which means that the symmetry-based indicator can only help us to diagnose the topological ``gapped states'', such as TI or TCI. 
For Na$_{3}$BrO, the system belongs to a spinless one with $\mathcal{T}$, which means that the symmetry-based indicator can only help us to diagnose the topological ``gapless states'', such as Weyl points, nodal lines or nodal rings. 
Similar diagnosing process can also be implemented in other class AII systems, which starts from a topological semimetal phase with violating the compatibility condition and ends with a trivial insulator/TI/TCI, and it will offer a way to predict/simulate topological phase transitions in materials.

\section{Conclusion}
The algorithm for diagnosing the complete topological information of systems violates compatibility condition and/or nontrivial symmetry-based indicator group condition is under the framework of symmetry-based indicators, but it compensates for the shortcomings of symmetry-based indicators. We display how to diagnose topological information in systems either violating compatibility condition or having a nontrivial symmetry-based indicator group condition with two material examples in class AI systems, i.e., spinless system with $\mathcal{T}$, to show both the effectiveness and correctness of this algorithm. We also discuss the limitation of the symmetry-based indicator, which cannot diagnose all the topological states in all cases. But we can still get the whole topological information by the compatibility condition. Furthermore, the algorithm is also discussed in class AII systems, i.e., spinful systems with $\mathcal{T}$ to show the topological phase transition between topological semimetals and topological (trivial) gapped states.  Similar discussion can also be made in other systems, like photons\cite{lu2014topological,lu2016topological,lu2015experimental,TOP_photonic4}, phonons\cite{zhang2018double,miao2018observation,Top_phon_mirco1,Top_phon_mirco2,Top_phon_mirco3,Top_phon_micro4,Top_phon_micro5,Weyl_phon_mecha,Top_acoustic,Weyl_acoustic4,Top_huber1,Top_HUBER2,Top_phonon6,liu2017model,xia2019symmetry}, and magnons\cite{li2017dirac,yao2018topological}.

\paragraph*{Acknowledgements}

We acknowledge the supports from JSPS KAKENHI Grants No. JP18H03678 and No. JP20H04633, Tokodai Institute for Element Strategy (TIES) funded by MEXT Elements Strategy Initiative to Form Core Research Center. T. Z. also acknowledge the support by Japan Society for the Promotion of Science (JSPS), KAKENHI Grant No. 21K13865

\bibliographystyle{unsrt}
\bibliography{reference}
 \newpage{}
\newpage{}

\end{document}